\documentclass{emulateapj}

\usepackage{amsmath}
\usepackage{amssymb}
\usepackage{graphicx}

\newcommand{\kms}{\ensuremath{\mathrm{km~s}^{-1}}}
\newcommand{\Mni}{\ensuremath{M_{\mathrm{Ni}}}}
\newcommand{\Mej}{\ensuremath{M_{\mathrm{ej}}}}
\newcommand{\Nifs}{\ensuremath{^{56}\mathrm{Ni}}}
\newcommand{\Cofs}{\ensuremath{^{56}\mathrm{Co}}}
\newcommand{\Msun}{{\ensuremath{\mathrm{M}_{\odot}}}}

\newcommand{\Min}{\ensuremath{M_{\rm i}}}
\newcommand{\Mf}{\ensuremath{M_{\rm f}}}

\newcommand{\tsn}{\ensuremath{t_{\rm sn}}}
\newcommand{\Lsn}{\ensuremath{L_{\rm sn}}}

\newcommand{\vph}{\ensuremath{v_{\rm ph}}}
\newcommand{\vsn}{\ensuremath{v_{\rm sn}}}
\newcommand{\Tph}{\ensuremath{T_{\rm ph}}}
\newcommand{\TI}{\ensuremath{T_{\rm I}}}
\newcommand{\vs}{\ensuremath{v_{\rm s}}}
\newcommand{\Xhe}{\ensuremath{X_{\rm he}}}

\shortauthors{Kasen \& Woosley}
\shorttitle{Type~IIP Supernova Light Curves}

\begin{document}

\title{Type~II Supernovae: Model Light Curves and Standard Candle Relationships}
\author{Daniel Kasen\altaffilmark{1,2}\email{kasen@ucolick.org} and S.E. Woosley\altaffilmark{1}}
\altaffiltext{1}{University of California, Santa Cruz}
\altaffiltext{2}{Hubble Fellow}

\begin{abstract} A survey of Type~II supernovae explosion models has
been carried out to determine how their light curves and spectra vary
with their mass, metallicity and explosion energy.  The presupernova
models are taken from a recent survey of massive stellar evolution at
solar metallicity supplemented by new calculations at sub-solar
metallicity. Explosions are simulated by the motion of a piston near
the edge of the iron core and the resulting light curves and spectra
are calculated using full multi-wavelength radiation transport.
Formulae are developed that describe approximately how the model
observables (light curve luminosity and duration) scale with the
progenitor mass, explosion energy, and radioactive nucleosynthesis.
Comparison with observational data shows that
the explosion energy of typical supernovae (as measured by kinetic energy at infinity)
varies by nearly an order of magnitude -- from $0.5$ to $4.0~\times
10^{51}$~ergs, with a typical value of $\sim 0.9 \times
10^{51}$~ergs. Despite the large variation, the models exhibit a tight
relationship between luminosity and expansion velocity, similar to
that previously employed empirically to make SNe~IIP standardized
candles.  This relation is explained by the simple behavior of
hydrogen recombination in the supernova envelope, but we find a
sensitivity to progenitor metallicity and mass that could lead to
systematic errors. Additional correlations between light curve
luminosity, duration, and color might enable the use of SNe~IIP to
obtain distances accurate to $\sim$20\% using only photometric data.
\end{abstract}


\section{Introduction}

Type II supernovae (SNe~II) result from the explosion of massive stars
that have retained their hydrogen envelope until their cores collapse
to neutron stars or black holes. The most common events, the Type~II
plateau supernovae, have a distinctive light curve, maintaining a
nearly constant luminosity for $\sim 100$~days, then suddenly dropping
off.  Upcoming synoptic surveys should discover millions of these events
out to redshifts of a few.  These observations will probe massive
stellar evolution in a broad range of galactic environments, and 
may also be used to measure cosmological distances and test
host galaxy dust properties.

SNe~II are diverse transients in terms of their luminosities,
durations, and expansion speeds. By modeling the observed light curves
and spectra,
one can constrain the physical
properties of the explosion such as the ejected mass, explosion
energy, and presupernova radius. In a pioneering study,
\cite{Hamuy_SNIIP} analyzed 
 16 observed supernovae and derived masses between $15$ and $50~\Msun$
and presupernova radii ranging from $70$ to $600~$R$_\odot$.  Unfortunately
these inferred values seem implausible, especially the high masses.  Direct
observations of the progenitors of nearby SNe~II from
pre-explosion images indicate masses of only $8-15~\Msun$
\citep{Smartt_prog}, while modern stellar evolution models predict
a mass range of $\sim 12-25$~\Msun\ \citep[e.g.,][]{Woo02,Woo07}.  The
overly large values inferred by \cite{Hamuy_SNIIP} likely reflect
deficiencies in the theoretical models used in the
light curve analysis \citep{LN_85}. More recent models tailored to individual events
have returned more reasonable numbers
\citep[e.g.][]{Utrobin_05cs, Baklanov}. 

Observations of SNe~II are also  used to measure cosmological distances.  In terms of brightness alone, they are poor standard
candles with luminosities varying by more than an order of magnitude,
but various methods can be used to standardized them.  Most previous
studies were of the Baade-Wessellink type, e.g., the expanding
photosphere method
\citep{Kirshner_EPM,Eastman_EPM,Dessart_EPM,Dessart_05cs,Jones_EPM}
and the spectral-fitting expanding atmosphere method 
\citep[SEAM,][]{Mitchell_1987A,Baron_99em}. Both approaches require detailed
atmospheric modeling. More recently, \cite{Hamuy_SCM} suggested a
much simpler approach using an empirical correlation between plateau
luminosity and expansion velocity (as measured from the Doppler
shift of spectral lines).  This standardized candle method has been
extended out to moderate redshifts with promising results \citep{Nugent_SCM,
Dovi_SCM}.  However, the physical underpinnings of the  method and its
potential vulnerability to systematic error have not yet been fully explored.

Here we present a new survey of SNe~IIP models based on the
one-dimensional explosion of realistic progenitor star models of
various masses, metallicities, and explosion energies.  We calculate
the broadband light curves and the detailed spectral evolution using a
code that includes a full multi-wavelength solution to the radiation
transport problem. The models allow us to explore the light curves
dependence on model parameters, which we compare to analytic
expectations.  We also reproduce and illuminate the physics leading to
the standardized candle relation.

\clearpage

\section{Analytical Scalings}
\label{sec:an}

Some insight into the light curves of SNe~IIP can be gained from
analytical scalings which express how the observables -- luminosity,
\Lsn, light curve duration, $\tsn$, and expansion velocity, \vsn --
depend on the basic supernova parameters -- the explosion energy, $E$,
ejecta mass, $\Mej$ and presupernova radius, $R_0$.  Although the analytic
light curves themselves are only approximate, the scaling relations
can be quite accurate \citep{Arnett_80,Chugai,Popov}.  As a guide to the
model results below, we rederive here some basic results using simple
physical arguments.

Although the mechanism of core-collapse supernova explosions is
complicated, the end result is the deposition of order $E \sim
10^{51}~{\rm ergs} = 1$~B near the center of the star, and
subsequently, the propagation of a blast wave that heats and ejects
the stellar envelope.  At the time the shock reaches the surface
(hours to days), the explosion energy is roughly equally divided
between internal and kinetic energy.  In the hydrogen envelope,
radiation energy dominates the thermal energy of ions and electrons by
several orders of magnitude.

In the subsequent expansion, the internal energy is mostly converted
into kinetic.  The ejecta is optically thick to electron scattering,
and the adiabatic condition gives $E_{\rm int}(t) = E_0
(R(t)/R_0)^{-1}$, where $E_0 \approx E/2$ is the initial internal
energy, and $R$, the radius.  After many doubling times, the ejecta
reaches a phase of free homologous expansion, where the velocity of a
fluid element is proportional to radius $R = vt$.  The final velocity
of the ejecta is of order
\begin{equation}
\vsn \approx (2 E/\Mej)^{1/2}
= 3\times 10^8 ~M_{10} E_{51}~{\rm cm~ s}^{-1},
\label{Eq:vel}
\end{equation}
where $M_{10} = \Mej/10~\Msun$, $E_{51} = E/10^{51}$~ergs.  For
homologous expansion, the internal energy evolves as $E_{\rm int}(t) =
E_0 (t/t_e)^{-1}$ where $t_e = R_0/\vsn$ is the expansion time.
Dimensionally, the luminosity of the light curve will then be
\begin{equation}
\Lsn = \frac{E_{\rm int}(\tsn)}{\tsn} = \frac{E_0 t_e}{\tsn^2},
\label{Eq:Lsn}
\end{equation}
where \tsn\ is the appropriate timescale for the duration of the light
curve.  We determine \Lsn\ and \tsn\ below for three different
scenarios. 

First, if the opacity, $\kappa$, in the supernova envelope were a
constant, the time scale of the light curve would be set by the
effective diffusion time.  Given the mean free path $\lambda_p =
(\kappa \rho)^{-1}$ and the optical depth of the ejecta $\tau =
R/\lambda_p$, the diffusion time is
\begin{equation}
\tsn = \tau^2 \frac{\lambda_p}{c} = \frac{R^2 \kappa \rho}{c}.
\label{Eq:tdiff}
\end{equation}
Over time, the supernova radius increases and the density decreases
due to the outward expansion.  Using the characteristic values
$R(\tsn) = \vsn \tsn$ and $\rho(\tsn) \sim \Mej/R(\tsn)^3$ in
Eq.~\ref{Eq:tdiff}, one can solve for \tsn.
\begin{equation}
\begin{split}
\tsn &\propto E^{-1/4} \Mej^{3/4} \kappa^{1/2}\\
\Lsn &\propto E~\Mej^{-1} R_0~\kappa^{-1},
\end{split}
\label{Eq:Arnett}
\end{equation}
where the scaling for \Lsn\ was determined by using \tsn\ in
Eq.~\ref{Eq:Lsn}.  These are the scalings of \cite{Arnett_80}

The results Eq.~\ref{Eq:Arnett}, though commonly applied, are not quite
adequate for SNe~IIP because the assumption of constant opacity
neglects the important effects of ionization \citep{Grassberg_1971}. Once the outer layers of
ejecta cool below $\TI \approx 6000$~K, hydrogen recombines
and the electron scattering opacity drops by several orders of
magnitude.  A sharp ionization front develops -- ionized material
inside the front is opaque, while neutral material above the front is
transparent.  The photon mean free path in the ionized matter
($\lambda_{\rm mfp} \sim 10^{10}$~cm) is much smaller than the ejecta
radius ($R = \vs t \approx 10^{15}$~cm), so the supernova photosphere
can be considered nearly coincident with the ionization front.

As radiation escapes and cools the photosphere, the ionization front
recedes inward in Lagrangian coordinates, in what is called a
recombination wave.  The progressive elimination of electron
scattering opacity allows for a more rapid release of the internal
energy.  Once the ionization front reaches the base of the hydrogen
envelope, and the internal energy has been largely depleted, the light
curve drops off sharply, ending the ``plateau''.  Any subsequent
luminosity must be powered by the decay of radioactive elements
synthesized in the explosion (the light curve ``tail'').

To account for hydrogen recombination while ignoring, for the moment,
radiative diffusion \citep[as in][]{Woo88,Chugai}, we use the fact that the
photosphere is fixed near the ionization temperature \TI\ and radiates
a luminosity $L_p$ given by
\begin{equation}
L_p = 4 \pi R^2 \sigma \TI^4 
\label{Eq:Stefans}
\end{equation}
The light curve timescale \tsn\ is then given by how long it takes the
photosphere to radiate away all of the (adiabatically degraded)
internal energy -- in other words, $L_p \tsn = E_0 (t_e/\tsn)$.  This,
(along with $R = v_s \tsn$) gives 
\begin{equation}
\begin{split}
\tsn &\propto E^{-1/8} \Mej^{3/8} R_0^{1/4} \TI^{-1},  \\
\Lsn &\propto E^{3/4} \Mej^{-1/4} R_0^{1/2} \TI^{2}.
\end{split}
\label{Eq:Chugai}
\end{equation}
Unlike Eqs.~\ref{Eq:Arnett}, these scalings show no dependence on
$\kappa$.  Effectively, the assumption is that the ejecta is infinitely
opaque below the recombination front, and fully transparent above.  We
will see in the numerical models that this assumption is not totally
correct, and hence diffusion below the photosphere is important.

Finally, to derive scalings which include both the effects of
radiative diffusion and recombination \citep[as in][]{Popov}, we return
to the diffusion time equation of Eq.~\ref{Eq:tdiff}, but now realize
that the radius of the opaque debris changes over time, not only due
to the outward expansion but also from the inward propagation of the
recombination front.  This photospheric radius is determined from
Eq.~\ref{Eq:Stefans}
\begin{equation}
R_i^2 = \frac{L_p}{4\pi \sigma T_i^4} = \frac{E_0 t_e}{\tsn^2 4\pi ,
\sigma T_i^4}
\end{equation}
where in the last equality we used Eq.~\ref{Eq:Lsn} to rewrite $L_p =
\Lsn$.  Plugging this expression for the radius into the numerator of
 Eq.~\ref{Eq:tdiff} gives
\begin{equation}
\begin{split}
\tsn &\propto E^{-1/6} \Mej^{1/2} R_0^{1/6} \kappa^{1/6} \TI^{-2/3} \\ 
\Lsn &\propto E^{5/6} \Mej^{-1/2} R_0^{2/3} \kappa^{-1/3} \TI^{4/3}.
\end{split}
\label{Eq:popov}
\end{equation}
These scalings are identical to those found by \cite{Popov} in a more
involved analysis. 

So far, the light curves described have only accounted for shock
deposited energy. Radioactive \Nifs\ synthesized in the explosion
introduces an additional energy source concentrated near the center of
the debris.  The heating from radioactive decay helps maintain the
ionization of the debris and so extends the duration of the plateau.
To incorporate this effect into the scalings, we generalize the
expression for the internal energy
\begin{equation}
E_{\rm int}(t) = E_0 \frac{t_e}{t}
  + E_{\rm ni}  \frac{t_{\rm ni}}{t}
  + E_{\rm co}  \frac{t_{\rm co}}{t},
\end{equation}
where $E_{\rm ni} \approx 0.6 \times 10^{50} \Mni$~ergs, $E_{\rm co} \approx
1.2 \times 10^{50} \Mni$~ergs are the total energy released from
\Nifs\ and \Cofs\ decay (with \Mni\ in units of \Msun) 
and $t_{\rm ni} \approx 8.8, t_{\rm co} \approx 113$~days are the
lifetimes. This correction for radioactivity can be written 
$E_{\rm int}(t) = E_0(t_e/\tsn) f_{\rm rad}$ with
\begin{equation}
f_{\rm rad} =  1 + 
0.26 \frac{\Mni}{E_{51} } \frac{t_{\rm co}}{t_e},
\label{Eq:niscale}
\end{equation}
where $E_{51} = E/10^{51}$~ergs.  Following the same arguments leading
to Eq.~\ref{Eq:popov}, we see that the plateau timescale scales as
$\tsn
\propto f_{\rm rad}^{1/6}$. For example, a \Nifs\ mass of $0.1~\Msun$
should extend the plateau by $\sim 24\%$ for $E_{51} = 1$.  Although one
might anticipate a change to \Lsn\ as well, the models show that
the decay energy does not typically have enough time to diffuse out
and affect the plateau luminosity (see Section \ref{sec:onemod}).

Below we compare the analytic scalings to our numerical simulations
and show that the simple relations, particularly
Eqs.~\ref{Eq:popov}, agree quite well.  The models allow us to refine
the exponents and calibrate the numerical constants in front.

\section{Progenitor Models}

\begin{deluxetable}{cccccc}
\tablewidth{0pt}
\tablecaption{Progenitor Star Models}
\tablehead{\colhead{$\Min$} & \colhead{$Z$} & \colhead{$\Mf$} & \colhead{$R_0$}
& \colhead{$M_{\rm Fe}$} & \colhead{$X_{He}$} }
\startdata
12 & 1.0 & 10.9 & 625  & 1.365 & 0.30\\
15 & 1.0 & 12.8 & 812  & 1.482 & 0.33\\
15 & 0.1 & 13.3 & 632  & 1.462 & 0.33\\
20 & 1.0 & 15.9 & 1044 & 1.540 & 0.38\\
25 & 1.0 & 15.8 & 1349 & 1.590 & 0.45\\
\enddata
\end{deluxetable}

\begin{figure}
\includegraphics[width=3.5in]{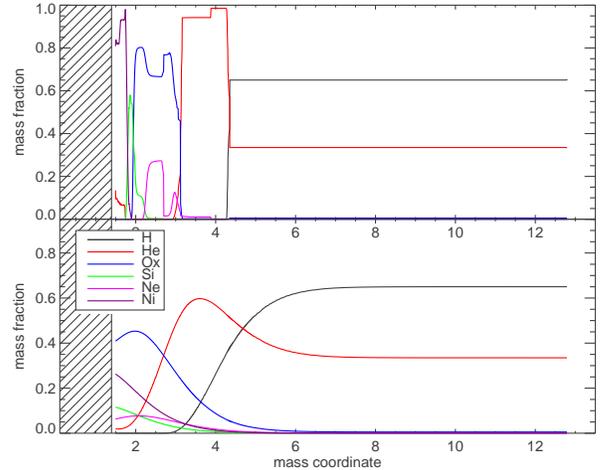}
\caption{Abundance distribution of model M15\_E.12 after explosion assuming 
either no mixing (top panel) or the mixing applied in this paper
(bottom panel).  The inner hashed region form the remnant..
\label{Fig:model_abun} }
\end{figure}

For our numerical models, we consider stellar progenitors with main
sequence masses in the range $\Min = 12-25~\Msun$, the range
expected to produce most of the observed events.  Properties of the
presupernova models are summarized in Table~1 which gives the zero age
main sequence and presupernova masses in solar masses ($\Min$ and
$\Mf$), the presupernova radius in solar radii ($R_0$), the iron core
mass in solar masses ($M_{\rm Fe}$), and the surface helium mass
fraction.  All models were computed with the Kepler code, which
follows stellar evolution including the most up-to-date opacities,
prescriptions for mass loss, and nuclear reaction rates
\citep{Rau02,Woo02,Woo07}.  Stars with larger initial masses
experience more mass loss, especially during the red giant phase,and
this narrows the range of final masses to $\Mf = 10.9-15.8$~\Msun.
For $\Min > 20~\Msun$, the presupernova mass declines with increasing
$\Min$.  More massive stars do, however, maintain significantly larger
radii at the time of explosion.

The helium mass fraction in the stellar envelope is also a function of
\Min\ (Table 1) varying from 30\% for $\Min = 12$~\Msun\ to 45\% for
$\Min = 25~\Msun$.  This variation is due to mass loss and convective
dredge up from the helium core, which are greater in more massive
stars.  As the envelope helium abundance affects both the electron
scattering opacity and the recombination temperature of the ejecta, we
will find it has a significant effect on the light curves of SNe~IIP.

While most stars in our survey have solar abundances, a lower
metallicity ($Z = 0.1$~solar) $15~\Msun$ model was also included.  The
chief effect of the lower metallicity was a smaller presupernova radius
and less total mass loss.

\section{Explosions and Nucleosynthesis}
\label{sec:explosion}

\subsection{Mass Cut, Fallback, and the  Production of \Nifs}

The explosion of each model was simulated by moving a piston outward
from an inner boundary at mass coordinate $M_{\rm pist}$
\citep{Woo95,Woo02}, typically taken to be the outer edge of the iron
core, and following the subsequent hydrodynamics assuming radial
symmetry. Each star was exploded several times to obtain variable
kinetic energies at infinity within the set of approximately $0.3,
0.6, 1.2, 2.4$, and $4.8 \times 10^{51}$ erg. The results are
summarized in Table 2. Polarization observations of SNe~IIP suggest
that the hydrogen envelopes are indeed spherically symmetric, although
the cores may appear aspherical, perhaps due to an asymmetric
explosion mechanism
\citep{Leonard_asph, Leonard_99em}.  Any asymmetry in the shock wave,
however, is likely smoothed out by propagating through the large
hydrogen envelope.

Explosive nucleosynthesis was calculated using the same code and
physics as in \citet{Woo07}. While the mass of \Nifs \ that is
synthesized is numerically well determined by this procedure, it is an
overestimate for two reasons. First, situating the piston at the edge
of the iron core, the deepest it can possibly be without violating
nucleosynthetic constraints on the iron isotopes, overestimates both
the density and mass close to the explosion. It is difficult to launch
a successful explosion in the face of such high accretion and a more
reasonable location for the piston might be farther out near the base
of the oxygen shell. There is a sudden increase in the entropy per
baryon, $S/N_Ak$, to a value around 4.0 signals a rapid fall off in
density in the presupernova star. To illustrate the sensitivity to the
piston location, a second version of the 0.6 B explosion of the solar
metallicity 15 \Msun \ star was calculated with the piston at the
location where $S/N_Ak$ = 4.0. The \Nifs \ production declined from
0.24
\Msun \ to 0.084 \Msun. 

Second, the ejection of \Nifs \ is sensitive to the treatment of
mixing and fallback. It was assumed here that whatever material had
positive speed at 10$^6$ s, the time of the link from the Kepler
hydrodynamics code to the spectral synthesis code would be
ejected. Since some of the slow moving material will fall back into
the collapsed remnant at later times, this also overestimates the
ejection of \Nifs. However, mixing which is largely finished before
$10^6$ s, reduces this sensitivity by taking \Nifs \ that would have
fallen in and moving it farther out in the ejecta \citep{Her94,Chu09}.

Observations of the tail of the supernova light curve constrain the
mass of \Nifs \ to be in the range $\sim$0.01 - 0.1 \Msun
\ \citep{Arn89,Smartt_prog} with values closer to 0.1 \Msun \ coming
from the more massive progenitors. All in all, it seems that
multiplying the \Nifs \ yield of our models by a factor of $0.25-0.5$
is reasonable. It should be noted that the plateau phase of the light
curve is insensitive to the \Nifs \ production (see Section \ref{sec:onemod}).

\subsection{Mixing}

In addition to its affect on the absolute yields, hydrodynamical
mixing during the explosion can carry \Nifs \ out into the hydrogen
envelope and hydrogen deep into the core of helium and heavy elements.
The early appearance of X-rays in SN 1987A and the smoothness of the
light curve showed that substantial mixing occurred - more than has
been provided so far in any calculation of just the Rayleigh-Taylor
instability. Mixing that commences with a broken symmetry in the
exploding core itself seems to be necessary \citep{Kif03}.  Because a
large number of models needed to be studied here and because the
degree of mixing is affected by the uncertain asymmetry of the central
engine, we used a simple parametric representation of the mixing
similar to that used by \citet{Pin88} and \citet{Heg09}.  A running
boxcar average of width $\Delta$M is moved through the star a total of
$n$ times until the desired mixing is obtained. The default values
$\Delta$M and $n$ are 10\% of the mass of the helium core and 4,
respectively. This gives, for example, the mixed composition for Model
15C in Figure~\ref{Fig:model_abun}.

We explored the effects of varying the degree of mixing, and found
that it lead to only small changes at the end of the plateau --
i.e. once the recombination wave had reached the inner layers of
helium and heavier elements.

\section{The Calculation of Light Curves and Spectra}
\label{sec:methods}

\begin{figure}
\includegraphics[width=3.3in]{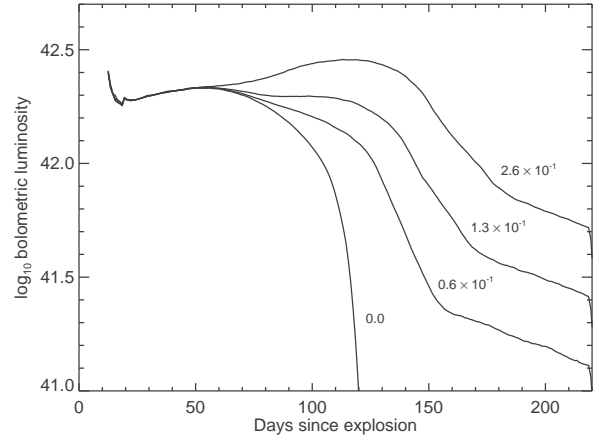}
\caption{Bolometric light curves of a  standard model (M15\_E1.2\_Z1) with 
different amounts of \Nifs\ ejected (marked on figure in units of
\Msun).  The radioactive energy deposition extends the plateau, but has 
little impact on the luminosity at $t < 50$~days.
\label{Fig:bol_gamma} }
\end{figure}

Several numerical studies of the light curves of SNe~IIP have 
been published \citep[e.g.,][]{LN_85,Young_SNII,Utrobin_99em,
  Nadyozhin_2003}.  One common limitation of the previous studies was that
the radiative transfer was often treated in the diffusion
approximation or with low wavelength resolution, although there have
been a few exceptions \citep[c.f.,][]{Baklanov, Chieffi_03}.  Non-LTE (NLTE)
radiative transfer calculations have been applied to the
stationary spectra of SNe~IIP \citep[e.g.,][]{Dessart_2005, Baron_93W}, but
not, so far, to time-dependent light curve calculations.

To calculate light curves and spectra of our models, we applied a novel method
which coupled a multi-wavelength implicit Monte Carlo radiation
transport code to a 1-dimensional hydrodynamics solver
\citep{Kasen_MC, Kasen_SBO}.  The initial conditions of the calculation were taken from
the Kepler explosion model at $t = 10^6$~s after explosion.  At
this time, the ejecta was largely homologous and the hydrodynamics
essentially unimportant.  While we therefore neglect the earliest part of
the light curve, our main interest here is the plateau phase.
Detailed radiative transfer calculations of the shock breakout phase
and early luminosity will be discussed in a separate paper
\citep{Kasen_SBO}.

In the Monte Carlo approach, the radiation field is represented by discrete 
photon packets which are tracked through randomized scatterings and absorptions.
At the start of the calculation, a large number $(\sim 10^5$) of packets were initiated in each zone.  The energy of the packets was chosen so that the sum equaled the
 equilibrium radiation energy of the zone. The initial frequency $\nu$ and direction vector $\hat{D}$ 
of each packet were sampled assuming that the distribution was isotropic and blackbody in
the comoving frame.   Throughout the simulation, additional packets were created to model 
gamma-rays input by the decay of \Nifs\ and \Cofs.  The transport and absorption of these gamma rays were likewise followed using a Monte Carlo approach applying the relevant opacities.

We adopted a mixed-frame approach for the transport whereby the gas
opacities and emissivities were calculated in the comoving frame, while
Monte Carlo packets were tracked in the lab frame. 
The relevant optical opacities included electron scattering, bound-free, free-free, and bound-bound line opacity, the last treated in the expansion opacity formalism of \cite{Eastman_93}.   The matter ionization and  excitation state were computed assuming Saha/Boltzmann statistics at
the matter temperature. While the code allows for
non-equilibrium between matter and radiation temperatures,
the radiation energy density is so dominant in SNe~IIP that the two equilibrated on a short timescale.

The scattering of photon packets was simulated by Lorentz transforming a packet into the comoving frame, preforming an isotropic scattering, and then transforming back to the lab frame.  The application of the two Lorentz transformations changes the energy and frequency of the outgoing packet. When averaged over many scattering events, this effect  accounts for the work done by the radiation field.  We checked that the correct behavior was recovered in very optical thick regions of the ejecta, where the radiation energy density evolved with
time as it should for a homologous adiabatic flow, $e_{rad} \propto (t/t_0)^{-4}$.

While the properties of individual packets were sampled from continuous distributions in space, time, and wavelength, the grid through which they moved was discrete.  In these calculations, the ejecta was divided into 150 equally spaced radial zones.  Opacities and emissivities in each zone were further defined on a wavelength grid of range $1-25,000$~\AA\ with a  constant binning of 5~\AA.    The physical properties of the zones  (e.g., density, temperature, ionization state, and opacity) were updated on a timescale chosen much  shorter than the dynamical timescale, with time-steps not exceeding $5 \times 10^4$~s.  Higher resolution tests were performed to confirm that the discretization was adequate. 

NLTE calculations of Type~II spectra show that deviations from LTE
have significant effects on line profiles, while the continuum flux is
less affected \citep{Baron_NLTE,Dessart_NLTET}.  To estimate the
potential effects, we computed stationary NLTE spectra on the plateau
(day 50 after explosion) using the same code.  A particularly relevant
NLTE effect is on the Ca~II IR-triplet, whose emission is
over predicted by LTE enough to cause a $\sim 0.1$~mag increase in
$I$-band magnitude.  While still relatively small, this error could be
significant when using the predicted $V-I$ color to correct for dust
extinction in observations.  We therefore used the NLTE spectral
results for the discussion in Section~\ref{sec:bc}.  Time-dependent NLTE
effects, which are not included here, can strongly affect the emission
in the hydrogen Balmer lines \citep{Dessart_NLTET} and the H-alpha
line in particular, which would modify the $R$-band magnitudes.

\section{A Typical Model}
\label{sec:onemod}

\subsection{Bolometric Light Curve}

\begin{figure}
\includegraphics[width=3.3in]{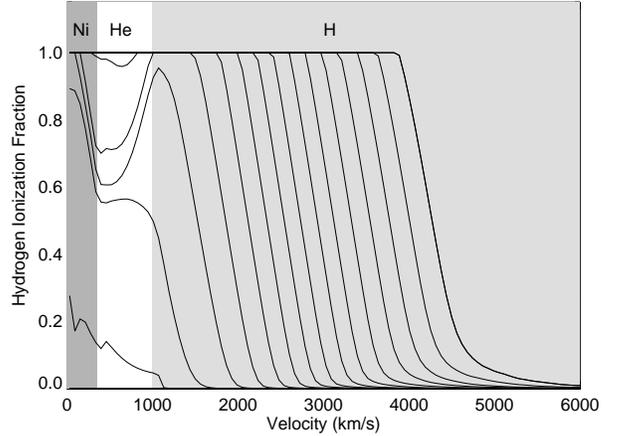}
\caption{Evolution of the hydrogen ionization fraction over time for 
model M15\_E1.2\_Z1. The ejecta photosphere forms near the ionization
front, which recedes inward in velocity (i.e., Lagrangian) coordinates
as the photosphere radiates and cools. The plateau ends when the
ionization front reaches the base of the hydrogen front.
\label{Fig:ion_wave} }
\end{figure}

\begin{figure}
\includegraphics[width=3.3in]{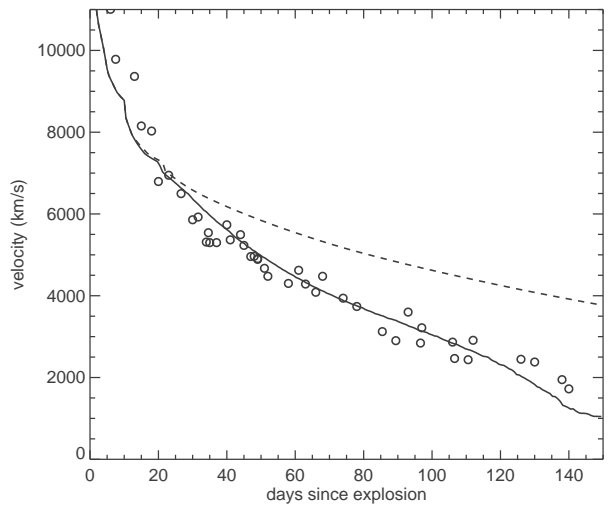}
\caption{Photospheric velocity evolution of model M15\_E1.2\_Z1 as a 
function of time.  The solid line shows the velocity of the electron
scattering photosphere defined at $\tau_e = 2/3$.  The dashed line
shows the photospheric velocity in a model in which hydrogen
recombination is neglected.  The circles are the observed velocities
of the sample of SNe~IIP compiled by \cite{Nugent_SCM}
\label{Fig:vrun} }
\end{figure}

We first focus on the properties of a typical SNe~IIP model
(M15\_E1.2\_Z1) which has parameters thought to be common: $\Min =
15~\Msun, E = 1.2$~B, and solar metallicity.
Figure~\ref{Fig:bol_gamma} shows the bolometric light curve for
different values of the ejected \Nifs\ mass.  Initially, the model luminosity
decreases after shock breakout, and reaches a minimum around day~20.  At
that time, the outermost layers of ejecta become cool enough such that
hydrogen can recombine.  This might be considered the beginning of the plateau phase.

On the plateau, the position of the photosphere is determined by the
location of the hydrogen recombination front, which occurs at a
temperature $\TI \approx 6000$~K.  As the ejecta cool, the front
recedes inward (Figure~\ref{Fig:ion_wave}).  At around
day 120, the front reaches the base of the hydrogen envelope. 
Recombination occurs more quickly in the helium layers, so the remaining
internal energy is depleted quickly and the light curve drops off
sharply, ending the plateau.

The evolution of the photospheric velocity over time (Figure~\ref{Fig:vrun})  agrees well with
the  observations of SNe~IIP compiled
by \cite{Nugent_SCM}.  Had we ignored the effects of hydrogen
recombination, the photospheric velocity would have declined much more
slowly, in conflict with the observed.  Thus, although the photospheric
velocity is often taken as a measure of $(E/\Mej)^{1/2}$ (i.e.,
Eq.~\ref{Eq:vel}) for SNe~IIP this is clearly only valid for times
before recombination sets in (here $t \la 20$~days). At later times,
the position of the photosphere is largely determined by the inward
progression of the recombination front.

\begin{figure}
\includegraphics[width=3.3in]{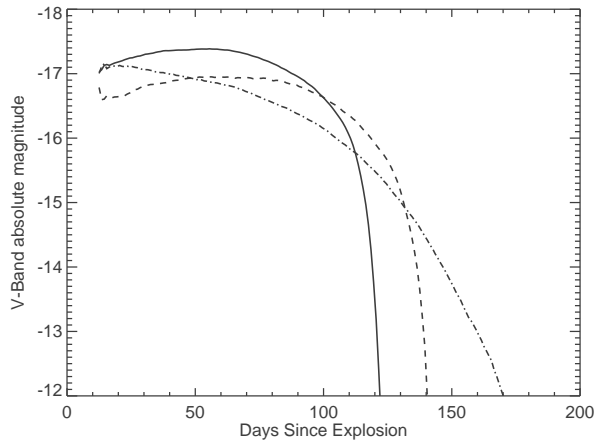}
\caption{V-band light curve of model M15\_E1.2\_Z1 (with no \Nifs\ included) 
computed using different opacity prescriptions. The solid line
properly includes all relevant opacities.  The dash-dotted line
shows a model in which hydrogen is not allowed to recombine, which
demonstrates the importance of ionization effects on the plateau.  The
dashed line shows a model in which the electron scattering opacity was
increased by a factor of 2, which demonstrates the importance of
diffusion in the ionized regions.
\label{Fig:opacity} }
\end{figure}

Figure~\ref{Fig:opacity} further illustrates how the nature of the opacity affects the light
curve. If we artificially increase the electron scattering opacity by a factor of
2, the light curve becomes dimmer and broader, in agreement with the
analytical scalings of Eq.~\ref{Eq:popov}.  This indicates that
radiative diffusion in the ionized regions is indeed significant.  If we
neglect the effects of hydrogen recombination, the resulting light
curve declines in a roughly power law fashion, with a lower average luminosity
and longer duration.  This implies that the recombination
wave is responsible for the flatness of the plateau and the steep drop
off afterward.

\begin{figure}
\includegraphics[width=3.3in]{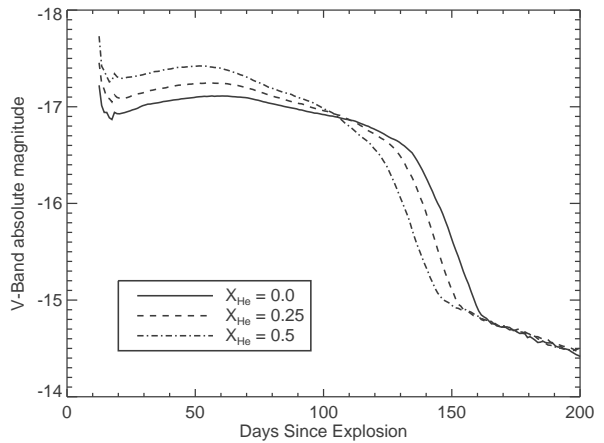}
\caption{Bolometric light curves of model M15\_E1.2\_Z1 (with a \Nifs\ mass 
of $0.06~\Msun$) computed with different helium abundances in the
hydrogen envelope.  A greater helium abundance reduces the electron
scattering opacity and leads to a shorter, brighter plateau.
\label{Fig:helium} }
\end{figure}

The opacity is also affected by the helium abundance \Xhe\ in the hydrogen
envelope.  Because helium recombines at higher
temperatures than hydrogen, a larger \Xhe\  effectively reduces the
electron scattering opacity.  The light curve of a model with $\Xhe = 0.5$ is therefore $0.4$~mag brighter and 20 days shorter
than one assuming pure hydrogen (Figure~\ref{Fig:helium}).  Helium in the core of the ejecta 
also affects the light curve, though in a slightly different way.  For models with larger helium cores, the recombination front will reach the base of the hydrogen layer at an earlier time,
and so the plateau will end relatively sooner.

 As expected from the analytical  arguments of
Section~\ref{sec:an}, the inclusion of radioactive \Nifs\ extends the plateau duration,
but has essentially no effect on the luminosity at times $\la 50$~days
(Figure~\ref{Fig:bol_gamma}).  Because \Nifs\ is synthesized only at
the ejecta center, radioactive energy does not have enough time to
diffuse out and affect the plateau unless extreme masses or outward
mixing of \Nifs\ are considered.

\subsection{Broadband Light Curves and Spectra}

\begin{figure}
\includegraphics[width=3.3in]{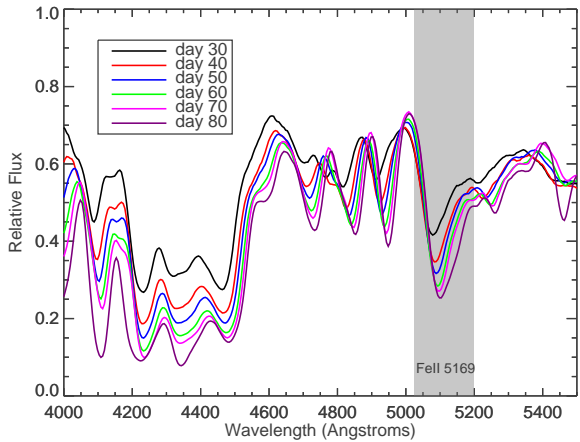}
\caption{Evolution of the Fe~II and Ti~II lines in the spectral range
4000-5550~\AA.  The recession of the photosphere is observable in the
decreasing Doppler shift of the line minima, especially that of
Fe~II~$\lambda 5169$ line.
\label{Fig:fe_vel} }
\end{figure}

\begin{figure}
\includegraphics[width=3.3in]{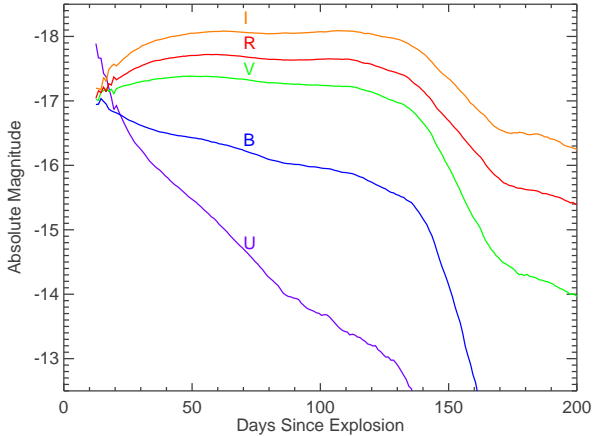}
\caption{Broadband UBVRI light curves
of our standard model M15\_E1.2\_Z1 (with an ejected \Nifs\ mass of
$0.13~\Msun$).  The faster decline in the U and B bands is due to
increasing line opacity (see Figure~\ref{Fig:spec_series}).
\label{Fig:broadband} }
\end{figure}

\begin{figure}
\includegraphics[width=3.3in]{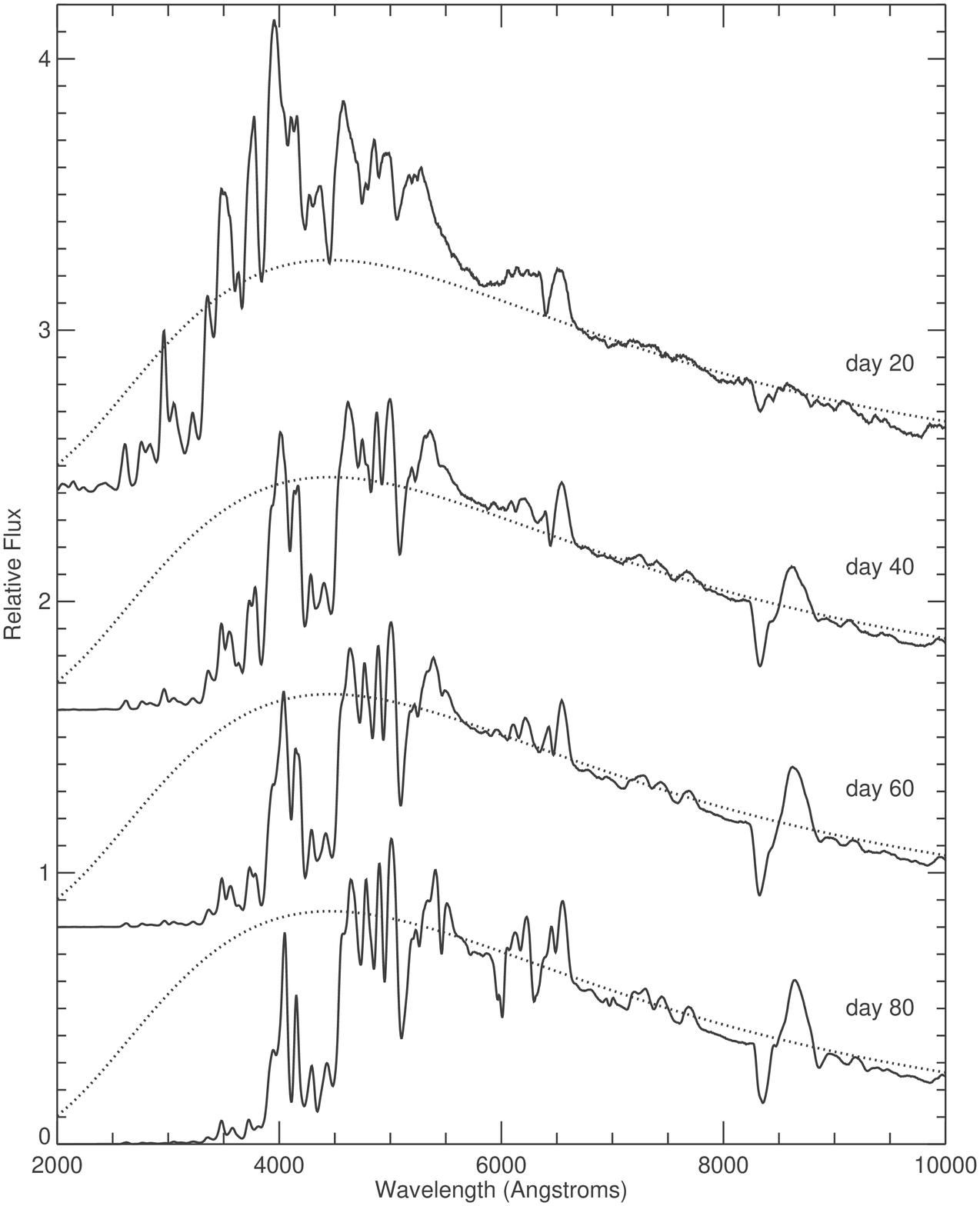}
\caption{Spectral evolution of model M15\_E1.2\_Z1, with the time
since explosion marked on the figure.  At longer wavelengths ($\lambda
\ga 5000~\AA$) the spectrum is well approximated by a blackbody at $T
\approx 6500$~K (dotted lines).  At shorter wavelengths, the iron group
line blanketing becomes progressively stronger over time, reducing the
flux.
\label{Fig:spec_series} }
\end{figure}

The model broadband light curves are shown in Figure~\ref{Fig:broadband}.  The $U$ and $B$ band light curves decline
sharply, showing virtually no plateau, while the $V$, $R$ and $I$ bands are
flatter.  This behavior, which is also seen in observations, can be understood by examining the spectral
evolution on the plateau (Figure~\ref{Fig:spec_series}).  At longer
wavelengths ($\lambda \ga 5000~\AA$) the continuum is fairly well
approximated by a blackbody of constant temperature.  The $V-I$ and
$V-R$ colors are therefore fairly constant over the plateau.  
At shorter wavelengths ($\lambda \la 5000~\AA$) on the other hand, the
spectrum is heavily affected by the blanketing from millions of
blended iron group lines (in particular those of Fe~II and Ti~II)
reflecting the metallicity of the progenitor star.  This line opacity
depends sensitively on temperature, as a slight cooling of the
photosphere induces Fe~III and Ti~III to recombine to Fe~II and Ti~II.
 The corresponding non-linear increase in line
blanketing, clearly visible in Figure~\ref{Fig:spec_series}, causes a
drop in $U$ and $B$ magnitudes much greater than would be expected from a
pure blackbody spectrum.

Figure~\ref{Fig:fe_vel} illustrates how the inward progression of the
supernova photosphere is detectable in the spectral series.  The
Doppler shifts of Fe~II and Ti~II absorption lines in the wavelength
region 4000-5000~\AA\ decrease over time.  In most applications, the
Fe~II~$\lambda 5169$ line is used to infer the photospheric velocity,
as it is strong enough to be measured relatively easily, but weak
enough to not be saturated above the photosphere.

\section{Model Survey}
\label{sec:survey}

\begin{deluxetable*}{ccccccccccc}
\tablewidth{0pt}
\tablecaption{Properties of Supernova Models}
\tablehead{
\colhead{Name} & 
\colhead{$M_i$} & 
\colhead{$E$}   & 
\colhead{$M_{\rm pist}$} &
\colhead{$\Mej$} &
\colhead{$\Mni$} &
\colhead{$L_{50}$}  &
\colhead{$t_{p,0}$}  &
\colhead{$M_{V,50}$}  &
\colhead{$v_{ph,50}$}
}
\startdata
M12\_E1.2\_Z1 &    12 &    1.21 &    1.36 &    9.53 &    0.16 &     1.91e42 &          116 &  -17.25 &         4915 \\
M12\_E2.4\_Z1 &      12 &    2.42 &    1.36 &    9.53 &    0.18 &     3.67e42 &           99 &  -17.98 &         6346 \\
M15\_E1.2\_Z1 &           15 &    1.21 &    1.48 &   11.29 &    0.26 &     2.16e42 &          124 &  -17.38 &         4959 \\
M15\_E2.4\_Z1 &           15 &    2.42 &    1.48 &   11.29 &    0.31 &     4.35e42 &          105 &  -18.15 &         6491 \\
M15\_E0.6\_Z1 &           15 &    0.66 &    1.48 &   11.25 &    0.24 &     1.26e42 &          149 &  -16.79 &         3966 \\
M15\_E4.8\_Z1 &           15 &    4.95 &    1.48 &   10.78 &    0.36 &     7.80e42 &           88 &  -18.80 &         8479 \\
M15\_E0.3\_Z1 &           15 &    0.33 &    1.48 &   11.27 &    0.22 &     5.93e41 &          177 &  -15.96 &         3125 \\
M20\_E1.2\_Z1 &           20 &    1.22 &    1.54 &   14.36 &    0.34 &     2.61e42 &          144 &  -17.57 &         4947 \\
M20\_E2.4\_Z1 &           20 &    2.42 &    1.54 &   14.37 &    0.40 &     4.85e42 &          119 &  -18.26 &         6459 \\
M20\_E0.6\_Z1 &           20 &    0.68 &    1.54 &   14.36 &    0.32 &     1.40e42 &          167 &  -16.89 &         3979 \\
M20\_E4.8\_Z1 &           20 &    4.99 &    1.54 &   14.37 &    0.48 &     8.57e42 &           99 &  -18.91 &         8337 \\
M25\_E1.2\_Z1 &           25 &    1.22 &    1.59 &   14.22 &    0.37 &     3.94e42 &          131 &  -18.00 &         5033 \\
M25\_E2.4\_Z1 &           25 &    2.43 &    1.59 &   14.22 &    0.43 &     6.66e42 &          107 &  -18.59 &         6483 \\
M25\_E0.6\_Z1 &           25 &    0.66 &    1.59 &   14.11 &    0.34 &     1.96e42 &          154 &  -17.23 &         4281 \\
M25\_E4.8\_Z1 &           25 &    5.00 &    1.59 &   12.97 &    0.56 &     1.10e43 &           86 &  -19.17 &         7948 \\
M15\_E1.2\_Z0.1 &           15 &    1.26 &    1.46 &   13.27 &    0.12 &     1.67e42 &          130 &  -17.04 &         4716 \\
M15\_E2.4\_Z0.1 &           15 &    2.48 &    1.46 &   13.24 &    0.16 &     3.08e42 &          107 &  -17.71 &         6098 \\
M15\_E0.6\_Z0.1 &           15 &    0.65 &    1.46 &   13.28 &    0.10 &     8.59e41 &          156 &  -16.32 &         3671 \\
M15\_E4.8\_Z0.1 &           15 &    4.90 &    1.46 &   13.18 &    0.20 &     5.31e42 &           88 &  -18.30 &         7670 \\
\enddata
\end{deluxetable*}

\begin{figure}
\includegraphics[width=3.3in]{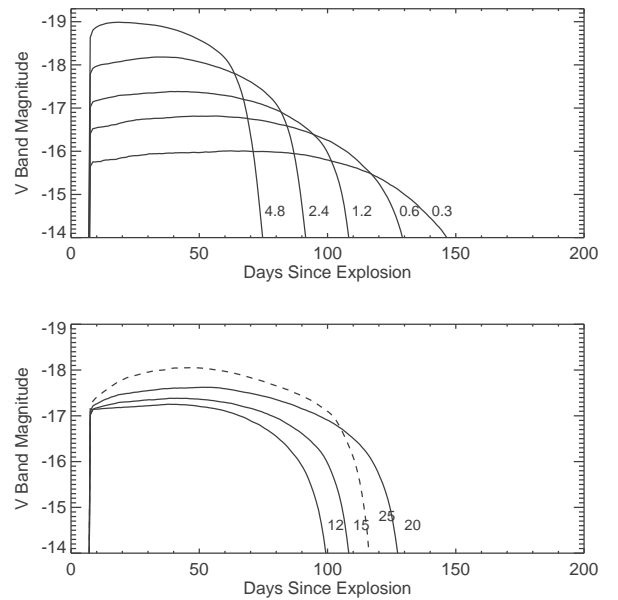}
\caption{Model $V$-band light curves from our survey, 
assuming no \Nifs ejected.  {\em Top:} light curves of the $M =
15~\Msun$ solar metallicity progenitor with different explosion
energies (marked on the figure in B).  Higher energy explosions are
brighter and shorter.  {\em Bottom:} light curves of 1.2~B explosions
with different progenitor initial masses (marked on the figure in
\Msun).  There is a non-monotonic behavior in the $M = 25~\Msun$ light
curve (dashed line) due to its increased mass loss and envelope helium
abundance.
\label{Fig:vbands} }
\end{figure}

\begin{figure}
\includegraphics[width=3.3in]{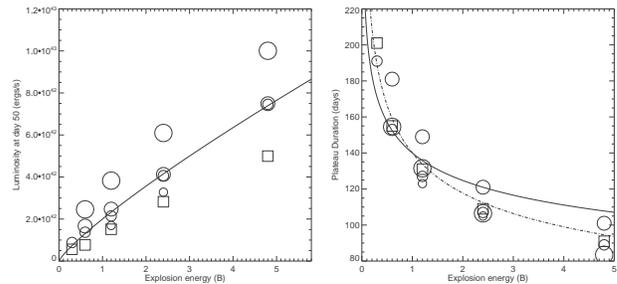}
\caption{Light curve properties of all of the models in our survey 
(assuming no \Nifs\ ejected).  {\em Left:} bolometric luminosity on
the plateau (measured 50 days after explosion) as a function of the
explosion energy.  Circles denote solar metallicity models, squares
$0.1$~solar models, and the size of the symbol is proportional to the
progenitor star initial mass. The solid line shows the $L
\propto E^{5/6}$ scaling. {\em Right:} same as the left panel, but for
the plateau duration. The solid line shows the $t_p \propto E^{1/6}$
scaling, the dashed line the $t_p \propto E^{1/4}$ scaling.
\label{Fig:en_lum} }
\end{figure}

Table~2 and Figures~\ref{Fig:vbands} and \ref{Fig:en_lum} summarize the light curve
properties of the entire model survey.  The models vary by more than a
factor of 10 in plateau luminosity, and by about a factor of 2 in
duration.  It is immediately clear that most of the variation in
SNe~IIP events reflects differences in explosion energy -- changes in
progenitor mass only account for a factor of $\sim 2$ in luminosity.
By directly comparing to the observed sample of nearby SNe~IIP (see
Section~\ref{sec:SC}), we infer that the explosion energy of real SNe~IIP
spans the range $0.6-4.8$~B, with a typical mean value around 0.9~B.

\begin{figure}
\includegraphics[width=3.3in]{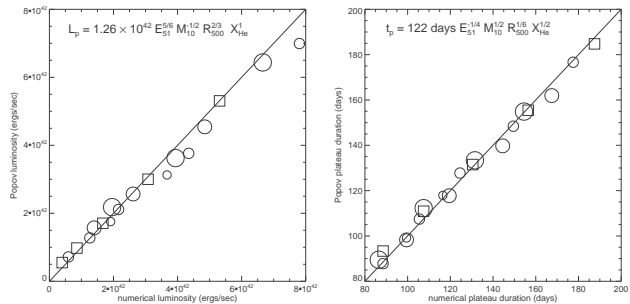}
\caption{Accuracy of the  analytic scalings Eqs.~\ref{Eq:popov}
for the plateau luminosity (left) and duration (right).  The $x$-axis
values are those determined by the numerical simulation, while the
$y$-axis values are those derived from the analytic equation. Circles denote
solar metallicity models, squares $0.1$~solar models, and the size of
the symbol is proportional to the progenitor star initial mass.
\label{Fig:popov} }
\end{figure}

Of the analytical scaling laws discussed in Section~\ref{sec:an}, the model
luminosity dependence follows most closely those of
Eqs.~\ref{Eq:popov}, and in particular $\Lsn \propto E^{5/6}$.  The
scaling of the plateau duration, however, deviates from
Eqs.~\ref{Eq:popov}, following more closely $t_p \propto E^{1/4}$.
Guided by the analytic results, we find expressions that well fit the
models
\begin{equation}
\begin{split}
L_{50}   &= 1.26 \times 10^{42}~ E_{51}^{5/6} M_{10}^{-1/2} R_{0,500}^{2/3} 
X_{\rm He}^{1}~{\rm ergs~s}^{-1}, \\
t_{p,0} &= 122~ E_{51}^{-1/4} M_{10}^{1/2} R_{0,500}^{1/6}
X_{\rm He}^{1/2}~{\rm days},
\end{split}
\label{Eq:fitscaling}
\end{equation}
where $R_{0,500} = R_0/500 R_\odot$ and $t_{p,0}$ is the plateau
duration when no \Nifs\ is included.  The \Xhe\ dependence
accounts for the effects of helium both in the envelope and the
core. Figure~\ref{Fig:popov} illustrates that the accuracy of these
expressions is quite good.

In principle, Eqs.~\ref{Eq:fitscaling} along with the expression for
the scaling velocity Eq.~\ref{Eq:vel}, could be used to infer the
physical parameters $(E,M,R_0)$ from the observed $(L,t_p,v)$, in the manner
applied by \cite{Hamuy_SNIIP}.  In practice, there are several
complicating factors.  The envelope helium abundance and the size of
the helium core, for instance, are significant factors, but
unfortunately there are no clean observables to constrain them. In
addition, the photospheric velocity on the plateau is largely
determined by recombination, and thus not necessarily a good measure
of the ejecta velocity $\vsn \propto (E/\Mej)^{1/2}$ (see
Figure~\ref{Fig:vrun}).  A measurement of the velocity at epochs prior
to recombination ($t \la 20$~days) is therefore preferred.

An alternative approach would use the fact that in the progenitor
models, $R_0$, $M_{ej}$, and $X_{\rm He}$ are correlated, so that some
of the degeneracies may be removed.  For future photometric surveys,
useful relations would allow for a determination of $E$ and $\Min$
given only $L_{50}$ and $t_{p,0}$.  We find for solar metallicity
models
\begin{equation}
\begin{split}
L_{50} &= 1.49\times10^{42}~ E_{51}^{0.82} M_{{\rm in},10}^{0.77}~ {\rm ergs} \\
t_{p,0}  &= 128~  E^{-0.26} M_{{\rm in},10}^{0.11}~ {\rm days},
\end{split}
\label{Eq:pscalings}
\end{equation}
where $M_{{\rm in},10} = \Min/10 \Msun$.  These relations (which fit
the models to within 10\%) can be applied to infer the gross
properties of SNe~IIP without need for follow-up spectroscopy.
However, one should bear in mind that they rely on the predictions of
stellar evolution and explosion calculations, and thus are subject to
uncertainties in, e.g., mass loss and fallback.

\begin{figure}
\includegraphics[width=3.3in]{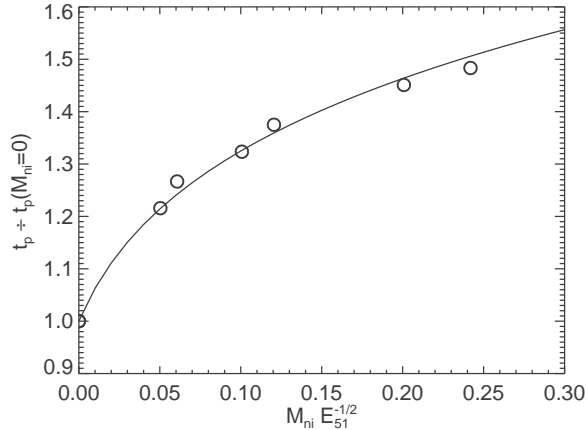}
\caption{Effect of radioactive \Nifs\ on extending 
the plateau duration.  The $y$-axis is the plateau duration $t_p$
divided by the $t_p$ for the zero \Nifs\ case.  The models included
here have $\Min = 15$~\Msun, explosion energies of 1.2 and 2.4~B, and
\Nifs\ mass varied from 0 to 0.35~\Msun.  The solid line shows the
analytic scaling Eq.~\ref{Eq:nislaw}.
\label{Fig:ni_tp} }
\end{figure}

Before applying either Eqs.~\ref{Eq:fitscaling} or
Eqs.~\ref{Eq:pscalings} it is critical to account for the fact that \Nifs\
in the ejecta tends to extend the plateau.  Figure~\ref{Fig:ni_tp}
shows that our derived analytical scaling (Eq.~\ref{Eq:niscale})  fits
reasonably well, with the refined numerical values
\begin{equation}
t_p = t_{p,0} \times 
\biggl(1 + 0.35 \Mni E_{51}^{-1/2} R_0^{-1}
\Mej^{1/2} \biggr)^{1/6}.
\label{Eq:nislaw}
\end{equation}
We find that the luminosity on the tail of the light curve is nearly
identical to the instantaneous energy deposition from \Cofs\ decay.
The ejected mass of
\Nifs\ can then  be inferred in the typical way,
 by measuring the luminosity at a point on the tail.  Unfortunately,
\Mej\ and $R_0$ also appear in this expression; however, their approximate
values for a given initial mass could be taken from Table~1.

\section{Bolometric Corrections and Dust}
\label{sec:bc}

\begin{figure}
\includegraphics[width=3.3in]{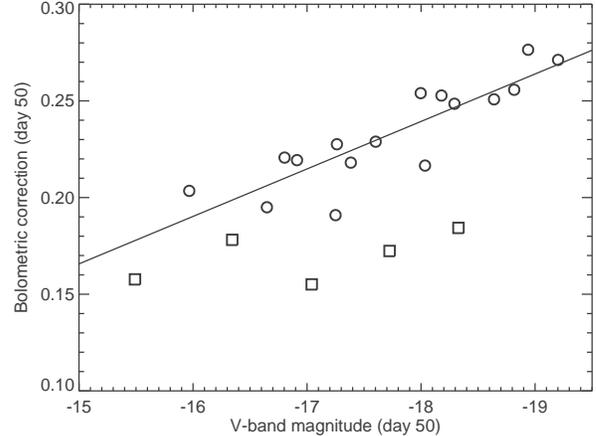}
\caption{Bolometric corrections (i.e., bolometric magnitude minus 
$V$-band magnitude) at day 50 for all models.  Circles are solar
metallicity models and squares $Z=0.1$~solar model.  The dashed line shows
a linear fit to the solar metallicity models, Eq.~\ref{Eq:bc}.
\label{Fig:bc} }
\end{figure}

From the models, one can derive formulae useful for making  bolometric and dust corrections to observations. 
Figure~\ref{Fig:bc} plots the difference in bolometric and $V$-band magnitude  at day
50 for all models. The typical bolometric corrections are around $0.2$~mag, but
increase for brighter events by as much as $0.07$~mag.  For solar
metallicity models, we fit the relation
\begin{equation}
BC_{50} = 0.24 - 0.025 \times (M_{V,50} + 18),
\label{Eq:bc}
\end{equation}
where $V_{50}$ is the $V$-band magnitude at day 50.  The $Z=0.1$~solar
metallicity models, due to the lesser line blocking, have bolometric
corrections about $0.07$~mag lower.  

\begin{figure}
\includegraphics[width=3.3in]{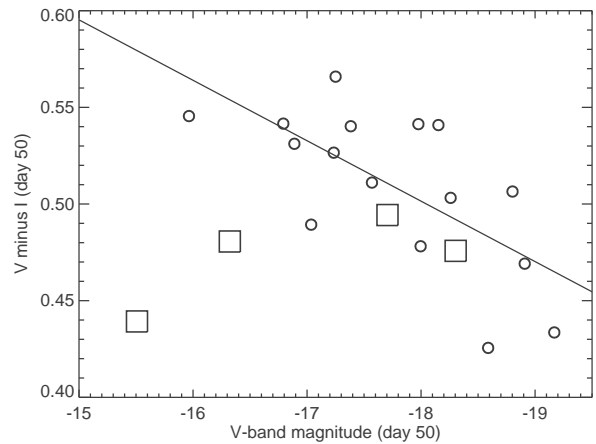}
\caption{The $V-I$ color at day 50 for all models.  Circles
denote solar metallicity models and squares 0.1 solar metallicity
models.  The solid line shows a linear fit to the solar metallicity
models, Eq.~\ref{Eq:vmi}.
\label{Fig:vmi}
}
\end{figure}

Previous studies have typically estimated dust extinction by measuring the $V-I$
excess over an assumed intrinsic color.  We find the models have a
roughly constant color on the plateau of $V - I \approx 0.5$; however,
there is a slight trend for brighter models to be bluer
(Figure~\ref{Fig:vmi}).  A fit to the solar metallicity models at day
50 gives
\begin{equation}
(V - I)_{50} = 0.52 + 0.03 \times (V_{50} + 17.5).
\label{Eq:vmi}
\end{equation}
The values are similar to the value $(V-I)_0 = 0.53$ that
\cite{Nugent_SCM} inferred by examining the ridge line of observed events.

As mentioned in Section~\ref{sec:methods}, the model $I$-band magnitudes are
sensitive to NLTE effects, especially in the Ca~II IR triplet line.
Eq.~\ref{Eq:vmi} was therefore determined using day 50 spectrum
calculations which treated calcium in NLTE, though under
the stationary approximation.  The model predictions are thus subject
to uncertainties in the assumed calcium abundance and perhaps to time
dependent NLTE effects \citep{Dessart_NLTET}.

\section{Standard Candle Relationship}
\label{sec:SC}

\begin{figure*}
\includegraphics[width=7in]{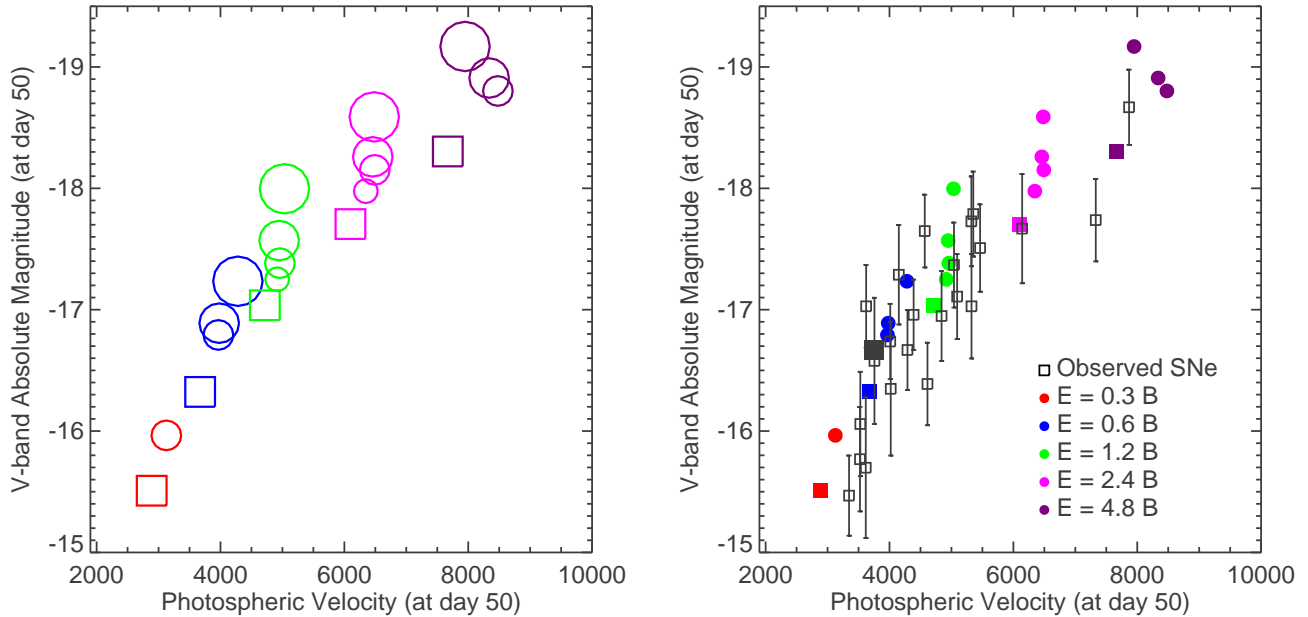}
\caption{Model standard candle relationship between 
the $V$-band plateau luminosity (measured at 50 days) and the
photospheric velocity.  {\em Left:} the models are color coded by
explosion energy.  Circles denote solar metallicity models, squares
$0.1$~solar models, and the size of the symbol is proportional to the
progenitor star mass.  {\em Right:} Comparison of the model relation
(circles) with the observation sample compiled by \citep{Hamuy_SNIIP}
(open squares).  The large filled square is SN~1999em using the
Cepheid distance of \cite{Leonard_ceph} and an extinction of $A_v =
0.31$.  \label{Fig:Hamuy} }
\end{figure*}

Using a sample of nearby observed SN~IIP, \cite{Hamuy_SCM} found that
the plateau luminosity (measured at day 50 after the explosion)
correlated rather tightly with the photospheric velocity, as measured
from the Doppler shift of spectral absorption lines.  This empirical
standard candle (SC) relation provides a simple means for calibrating
SN~IIP luminosity for distance measures.

Figure~\ref{Fig:Hamuy} shows the Hamuy SC relation for our model
survey set, here in terms of the $V$-band magnitude ($M_V$) and the
photospheric velocity.  The model relation is as tight or tighter than
the observed, and with a similar slope.  The rms dispersion is only
$\sigma = 0.27$, which translates to $\sim 13\%$ errors in distance
measures. To first order, the velocity and luminosity of SNe~IIP are both set by the
explosion energy.   The dispersion in the relation is due to
variations in the progenitor
mass and metallicity for a given explosion energy. 

The physical interpretation of the model SC relation is straightforward,
being essentially a recasting of the Baade-Wesselink or expanding
photosphere methods that have been in use for many years.  The
luminosity is written using Stefan's law and the radius of the
supernova photosphere $R = v_{\rm ph} t$
\begin{equation}
L = 4 \pi \vph^2 t^2 \zeta^2 \Tph^4,
\label{Eq:Lstef}
\end{equation}
where $\zeta$ is a ``dilution'' factor which accounts for deviation of
the spectrum from blackbody.  In Type~II atmospheres, both the
effects of scattering and line blanketing contribute to $\zeta$
\citep{Wagoner_EPM,Eastman_EPM}.  
To determine $L$ using the expanding photosphere method, the observer
measures \vph\ and the time since explosion $t$, and estimates the
photospheric temperature \Tph\ from the color of the spectrum.  The
dilution factor must be calculated using detailed numerical models
(the main complexity of the approach).  NLTE spectral modeling finds
that $\zeta$ varies between 0.5 and 2.0, and is chiefly a function of
luminosity, being rather insensitive to other ejecta parameters such
as the density structure
\citep{Eastman_EPM, Dessart_EPM}.

The standard candle relation is simply an expression of
Eq.~\ref{Eq:Lstef} under certain restricted conditions. The time since
explosion $t$ is, by construction, fixed at 50~days.  The temperature
\Tph\, for SNe~IIP on the plateau is nearly a constant, constrained to
be near the recombination temperature $T_i \approx 6000$~K.  The
dilution factor $\zeta$ may vary from event to event, but if $\zeta$
is primarily a function of luminosity this dependence can be absorbed into
the exponent.  This implies $L = C \vph^{2+\epsilon}$, where the
constant $C$ and the non-blackbody effects $\epsilon$ can be
calibrated using a sample of nearby objects, or a set of theoretical
models.

The SC relation need not be applied only at day 50, and we find that
similar relations
apply all along the plateau. However the time since explosion must be
known as the normalization depends on time (Eq.~\ref{Eq:Lstef}).  We
find that an uncertainty in explosion time of 10 days leads to an
error in inferred brightness of $0.2-0.3$~mag. It is unwise to apply the
SC relation at times much earlier than 30 days, as the ejecta
temperatures are likely too high for recombination to have set in, and
there is no assurance that $\Tph \approx T_i$. 

One nice feature of the models is that they offer an absolute
normalization of the SC relation without needing to assume a value of
the Hubble constant.  By fitting the relation evaluated at different
times since explosion, we find
\begin{equation}
\begin{split}
M_V(t) = -17.4 - 6.9 \log_{10} (v_{\rm ph}(t)/5000.0) 
\\ + 3.1 \log_{10} (t/50~{\rm days})
\label{Eq:SCR}
\end{split}
\end{equation}
The models do predict a deviation from the simple $L \propto \vph^2$
relation of Eq.~\ref{Eq:Lstef}, showing instead $L \propto
\vph^{2.75}$ in general accordance with that found in the observational sample
\citep{Hamuy_SCM}.  This effect is primarily due to the deviation of the spectrum from a 
blackbody.

The model relation of Figure~\ref{Fig:Hamuy} has a similar
normalization to the observations, taken from \cite{Hamuy_SNIIP}. This implies
 that our model SC relation is in rough agreement with the distances to
SNe~IIP obtained in other ways.  Particularly comforting is the
agreement with SN~1999em, which has a measured Cepheid distance to its
host galaxy NGC~1637 of $11.7 \pm 1.0$~Mpc \citep{Leonard_ceph}.  We
find a very similar distance of $11.6 \pm 1.2$~Mpc from
Eq.~\ref{Eq:SCR} when taking the observed values $m_v = 13.98$,
$\vph = 3757~\kms$, and \citep[following][]
{Baron_2000, Hamuy_EPM}, an extinction of $A_v = 0.31$.
This distance is also consistent with independent estimates using the expanding
photosphere method \citep{Dessart_EPM} and SEAM \citep{Baron_99em}.

\begin{figure}
\includegraphics[width=3.3in]{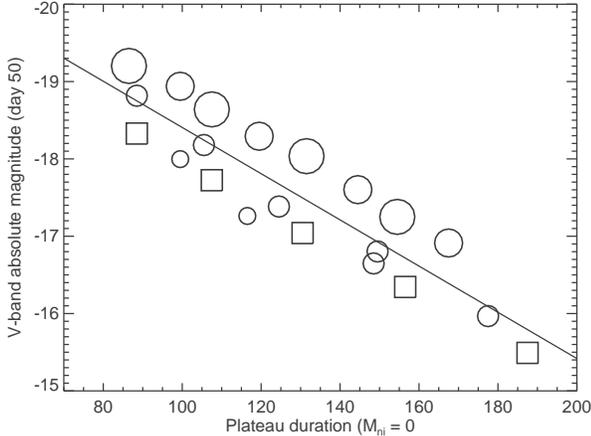}
\caption{Relationship between the plateau duration (assuming zero \Nifs) and
the luminosity at day 50.  Circles denote solar metallicity models,
squares $0.1$~solar models, and the size of the symbol is proportional
to the progenitor star mass.  
\label{Fig:ltp} }
\end{figure}

One drawback of the standard candle method, from the observational
point of view, is that a high quality spectrum is needed to measure
the photospheric velocity -- a difficult prospect for high redshift
events.  As future surveys will observe light curves for a enormous
number of SNe~IIP with limited spectroscopic follow-up, methods of
purely photometric calibration, however coarse, may be of interest.
As the explosion energy is the primary variable determining both the
plateau luminosity and duration, we explored the relationship between
these two observables.  A relationship exists (Figure~\ref{Fig:ltp})
and is fit by
\begin{equation}
M_{V,50} = -18.4 - 0.03 [ t_{p,0} - 100 ]
\end{equation}
Applying this relation reduces the dispersion from 1~mag down to
0.4~mag.  In practice,
the measured plateau duration $t_p$ must be corrected for the effect of 
the ejected \Nifs\ mass on its duration in order to determine $t_{p,0}$.   
The residual scatter in the relation is clearly due to variation in progenitor
initial mass or metallicity for a given explosion energy.  Presumably,
the scatter could be reduced further by using additional light curve relation,
such as the color evolution.

\section{Discussion and Conclusions}

We explored the light curves and spectra of SNe~II models with various
progenitor masses, metallicities, and explosion energies.  We found
that explosions with energies $0.3-4.8$~B of stars with initial masses
in the range $12-25~\Msun$ can explain the observed range of
luminosities, velocities, and light curve durations of most SNe~IIP.
For existing and future observational surveys, the model results
should be useful for inferring the progenitor star properties,
explosion energies, distances, and dust extinction of observed events.

This study, as have previous studies, quantified how the basic
supernova parameters (\Mej, $R_0$, and $E$) affect the light curves.
We also highlighted the important role of two additional parameters:
the radioactive \Nifs\ mass and the envelope helium abundance. The
presence of \Nifs\ extends the plateau duration, but typically does
not affect the luminosity on the plateau at times $t \la 50$~days.
The neglect of the effect of \Nifs\ may be the main reason why
\cite{Hamuy_SNIIP}, in his analysis of 16 SNe~IIP, 
inferred implausibly large ejecta masses (up to
$50~\Msun$).  In that study,
the longer plateau duration would have to be accounted for by an
increased diffusion time, and hence larger ejecta mass.  Here we
presented analytical formulae which may be useful in accounting for
the effects of \Nifs\ on the plateau.

The models confirm the standard candle method of calibrating SNe~IIP
and illuminate its physical origin.  The method is
a promising way to determine distances to SNe~IIP, with a clear
physical explanation in terms of the ionization physics of hydrogen.
On the other hand, the models raise some concerns about systematic
errors.  Progenitors with different masses or metallicities lie on
differently normalized relations in Figure~\ref{Fig:Hamuy}.  If the
progenitor population at high redshift  has different demographics 
than that at low redshift (as might be expected) a systematic bias may
be introduced into distance measurements.  The effect of going from
$Z=1$ to $Z=0.1$ solar metallicity in the models is at the 0.1~mag
level. It may be possible to reduce these errors by using color
information from the light curve.

We find a correlation between plateau luminosity and plateau duration
which could be useful in roughly calibrating SNe~IIP luminosities
using only photometric data (to about 20\% in distance).  This
correlation reflects the fact that in the models one parameter, the explosion
energy, primarily controls both the light curve brightness and duration, while the
progenitor star properties play a secondary role.  The validity of
such a relation needs to be empirically  checked, as the scatter will
be smaller or larger depending on whether the bulk of SNe~IIP arise
from a narrower or wider range of progenitor masses and radii than
that considered here.  In practice, the relation also  needs to take into account
the effect of \Nifs\ on extending the plateau duration.

Correction for dust extinction remains a difficult issue for
determining the distances to SNe~IIP.  The models provide some
theoretical guidance as to the intrinsic color evolution of SNe-IIP
light curves, however their accuracy may be limited by the assumptions
in the radiative transfer, and are sensitive to variations in the
envelope metallicity. On the other hand, one could try to invert the
problem.  Assuming the cosmological parameter are accurately
constrained by other means, one could use the standard candle method
to solve for the dust extinction of SNe~IIP, thus providing an estimate of
the variation of dust properties with galactic environment and
redshift.

\acknowledgements

The authors gratefully acknowledges helpful conversations on the
subject of the paper with Alex Heger, Peter Nugent, and Lars Bildsten. Support for DNK was provided by
NASA through Hubble fellowship grant
\#HST-HF-01208.01-A awarded by the Space Telescope Science Institute,
which is operated by the Association of Universities for Research in
Astronomy, Inc., for NASA, under contract NAS 5-26555.  This research
has been supported by the NASA Theory Program (NNG05GG08G) and the DOE
SciDAC Program (DE-FC02-06ER41438).  We are grateful for computer time
provided by ORNL through an INCITE award and at NERSC.

\end{document}